\documentclass[12pt]{iopart}
\expandafter\let\csname equation*\endcsname\relax
\expandafter\let\csname endequation*\endcsname\relax
\usepackage{iopams}
\usepackage{amsmath}
\usepackage{bm}
\usepackage{graphicx}
\usepackage{amssymb}
\usepackage{color,soul}
\usepackage[T1]{fontenc}
\usepackage{ae,aecompl}

\begin{document}

\title[A simple test for ideal memristors]{A simple test for ideal memristors}

\author{Y.~V.~Pershin}
\ead{pershin@physics.sc.edu}
\address{Department of Physics and Astronomy, University of South Carolina, Columbia, South Carolina
29208, USA}

\author{M.~Di Ventra}
\ead{diventra@physics.ucsd.edu}
\address{Department of Physics, University of California San Diego, La Jolla, California 92093, USA}

\vspace{10pt}
\begin{indented}
\item[]August 2017
\end{indented}

\begin{abstract}
An {\it ideal} memristor is defined as a resistor with memory that, when subject to a time-dependent current, $I(t)$, its resistance $R_M(q)$ depends {\it only} on the charge $q$ that has flowed through it, so that its voltage response is $V(t)=R_M(q)I(t)$. It has been argued that a clear fingerprint of these ideal memristors is a pinched hysteresis loop in their I-V curves. However, a pinched I-V hysteresis loop is {\it not} a definitive test of whether a resistor with memory is truly an ideal memristor because such a property is shared also by other resistors whose memory depends on additional internal state variables, other than the charge.
Here, we introduce a very simple and {\it unambiguous} test that can be utilized to check experimentally if a resistor with memory is indeed an ideal memristor. Our test is based on the {\it duality} property of a capacitor-memristor circuit whereby, for any initial resistance states of the memristor and any form of the applied voltage, the final state of an ideal memristor {\it must} be identical to its initial state, if the capacitor charge finally returns to its initial value. In actual experiments, a sufficiently wide range of voltage amplitudes and initial states are enough to perform the test. The proposed test can help resolve some long-standing controversies still existing in the literature about whether an ideal memristor does actually exist or it is a purely mathematical concept. % {\bf We bet that it does not!}
\end{abstract}

\maketitle
Although materials and systems that show a resistive response with memory or, in other words, time non-local resistance (see Ref.~\cite{pershin11a} for a thorough review) have been known for about two centuries~\cite{prodromakis2012two}, in 1971 a particular class of two-terminal resistive devices with memory has been suggested~\cite{chua71a}. These devices have been postulated to satisfy the following equation~\cite{chua71a}
\begin{equation}\label{eq:1}
  V_M(t)=R_M(q)I(t),
\end{equation}
where $V_M$ and $I$  are the voltage across and current through
the particular device, and $R_M(q)$ is its resistance which depends {\it only} on the charge $q$ that has flowed through the
device from an initial moment of time. These devices have been given the name of ``memristors''~\footnote{The definition of flux-controlled memristors is equivalent to Eq. (\ref{eq:1}).}.

Note that, although the name ``memristor'' is nowadays used liberally to indicate any resistive device with memory, the mathematical definition~(\ref{eq:1}) is quite strict. In fact,
it can be shown, using the theory of response functions~\cite{kubo1957statistical}, that {\it any} resistive element subject to a time-dependent voltage (or current) can be described by
the set of equations~\cite{di2013physical}
\begin{eqnarray}
I(t)&=&R^{-1}\left(x,V_M,t \right) V_M(t) \label{Geq1}\\ \dot{x}&=&f\left(x,V_M,t\right) \label{Geq2}
\end{eqnarray}
where $f$ is some vector function of internal state variables, $x$, other than the charge. These are the variables that provide memory to the system, and come in various physical realizations, such as spin polarization,
atomic position of defects, etc.~\cite{pershin11a}.

Devices that satisfy Eqs.~(\ref{Geq1}) and~(\ref{Geq2}) are more aptly called ``memristive''~\cite{chua76a} to distinguish them from {\it ideal memristors} as those represented
by Eq.~(\ref{eq:1}). Of course, from a purely mathematical point of view, the devices that are described by Eq.~(\ref{eq:1}) can be viewed as a subset of those described by Eqs.~(\ref{Geq1}) and~(\ref{Geq2}), by simply stating that the {\it only} state variable is the charge that flows through the device: $x=q$ and $f$ is
then simply  the current $I(t)=\textnormal{d}q/\textnormal{d}t$.

However, this seemingly innocuous mathematical reduction hides a very important {\it physical} question: do devices that are represented by Eq.~(\ref{eq:1}) (ideal
memristors) actually exist in Nature?

\begin{figure}[bt]
	\centering{\includegraphics[width=70mm]{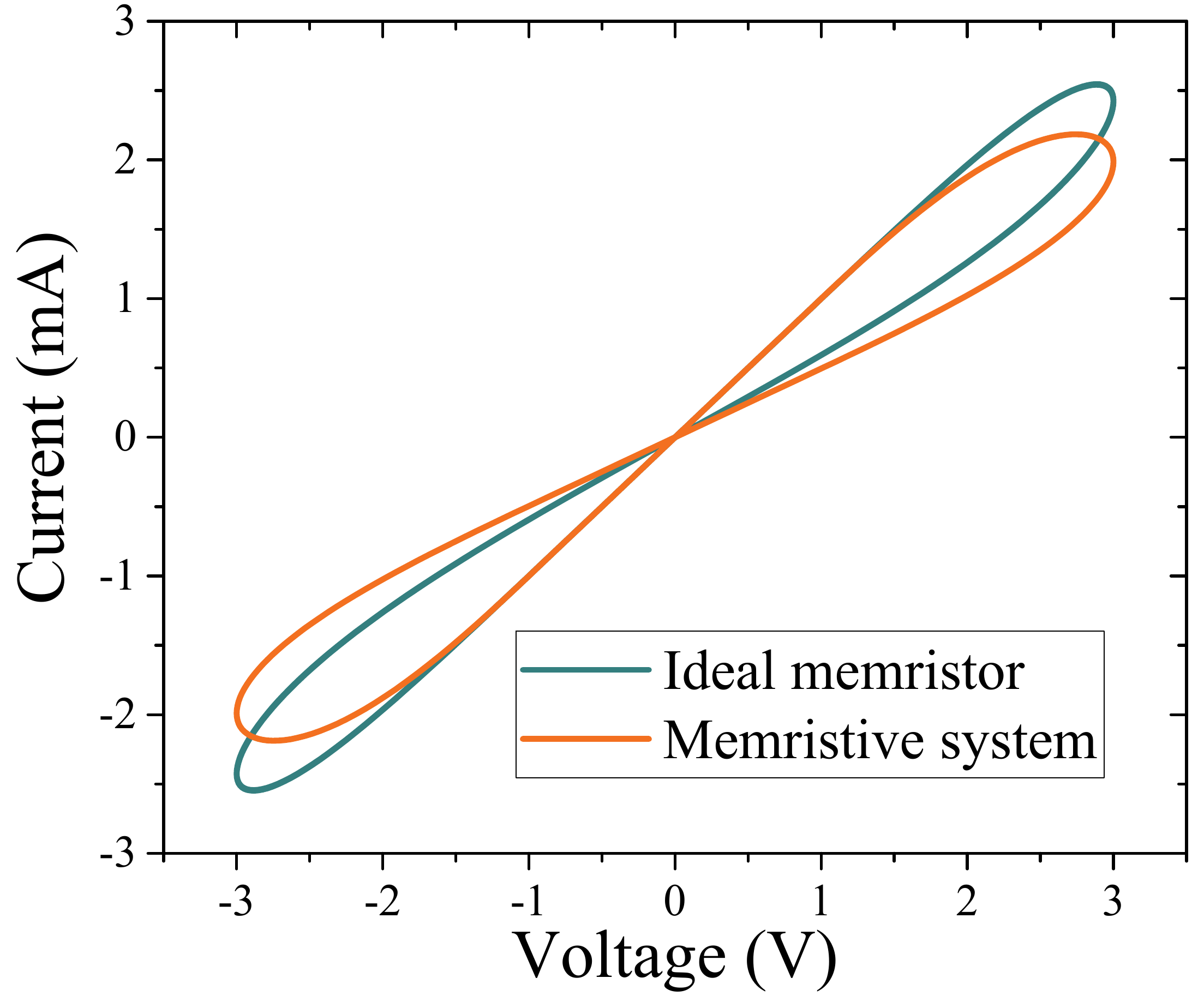}}
	\caption{Current-voltage characteristics of the ideal memristor (Eq. (\ref{eq:2}) as a model for Eq. (\ref{eq:1})) and a threshold-type memristive system (Eqs. (\ref{eq:3}) and~(\ref{eq:4}) as models for Eqs. (\ref{Geq1}) and~(\ref{Geq2}), respectively) driven by a sinusoidal voltage source $V(t)=V_0\sin (2\pi\nu t)$.
		The plot has been obtained by using $V_0=3$~V, $\nu=0.5$~kHz, $R_M(t=0)=R_{on}=R_0=1$~k$\Omega$,  $R_{off}=10R_{on}$, $\alpha=3\cdot10^{11}$~k$\Omega/$C$^2$, $\beta=10^3$~k$\Omega$/V$\cdot$s, $V_t=1$~V.}
	\label{fig:1}
\end{figure}

Researchers claiming that ideal memristors actually do exist point to the well-known consequence of Eq.~(\ref{eq:1}) that the corresponding $I$-$V$ curves are pinched hysteresis loops, and this has been claimed to serve
as {\it the} fingerprint of ideal memristors~\cite{Adhikari13a,chua2014if} (see the ideal memristor curve in Fig.~\ref{fig:1}). This argument, however, is very weak and not definitive since also other resistive devices whose memory depends on internal state variables other than the charge, do show pinched hysteresis loops (see, e.g., Ref.~\cite{pershin11a} and the memristive system curve in Fig.~\ref{fig:1}).

This fact has contributed to debates in the literature~\cite{mouttet2012memresistors,vongehr2015missing} concerning the memristor ``discovery''~\cite{strukov08a}. In addition, from the physical point of view, the present authors~\cite{di2013physical} and others~\cite{meuffels2012fundamental,sundqvist2017memristor}  have identified several drawbacks of the definition~(\ref{eq:1}) as a possible physically realizable device. These drawbacks include incompatibility with the symmetries of Maxwell's equations~\cite{di2013physical}, violation of Landauer's principle of minimal heat generation~\cite{meuffels2012fundamental,di2013physical}, and stochastic catastrophe~\cite{di2013physical}.

\begin{figure}[b]
\centering{\includegraphics[width=70mm]{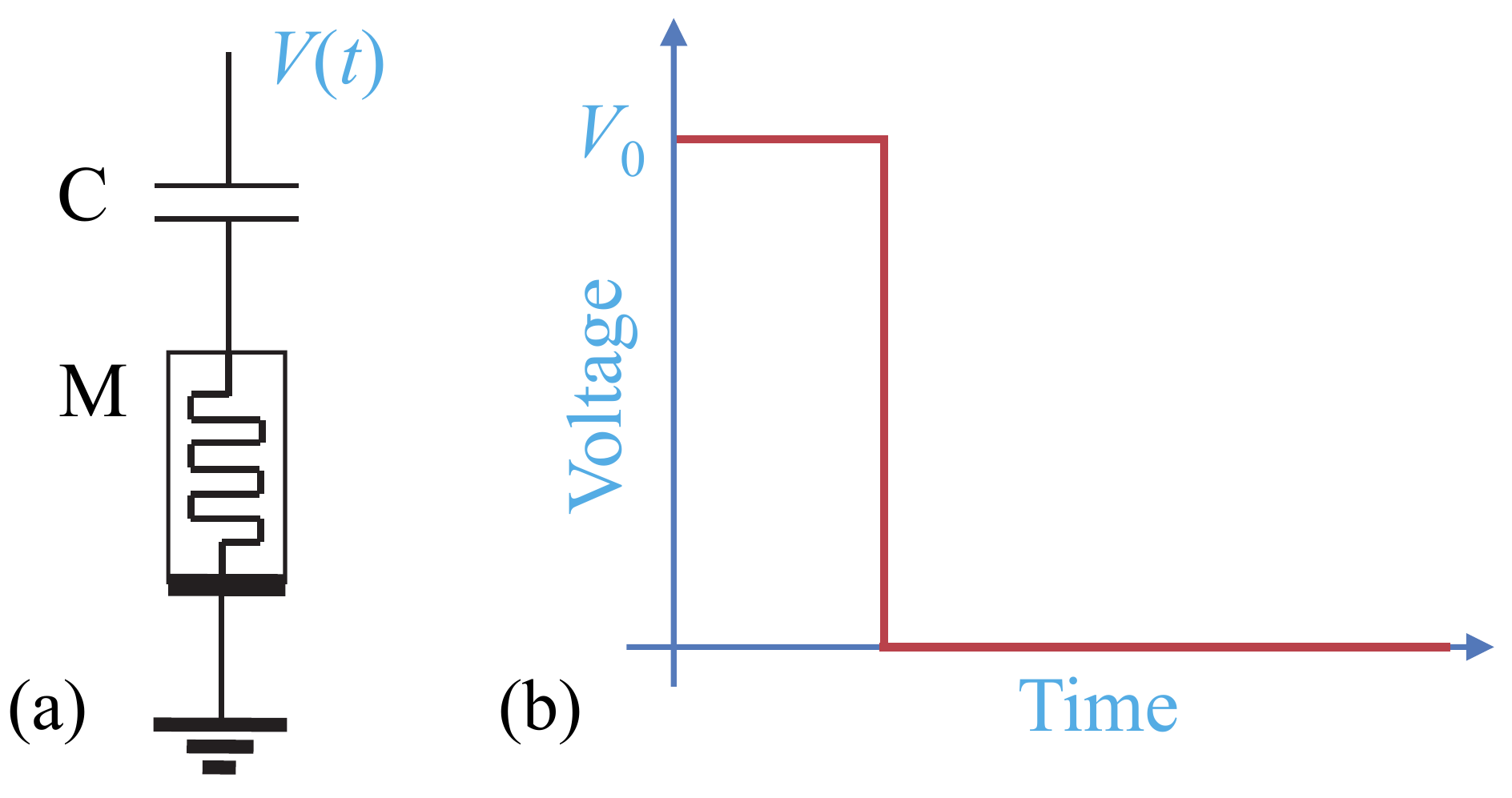}}
\caption{(a) Capacitor-memristor circuit used for the suggested memristor test. (b) The simplest form of the testing voltage $V(t)$.}
\label{fig:2}
\end{figure}

Since the above  physical flaws of a possible device represented by Eq.~(\ref{eq:1}) do not seem to have settled the question posed above, in this letter we introduce a very simple test that can be done in any laboratory to {\it unambiguously} distinguish ideal memristors (Eq.~(\ref{eq:1})) from all other resistive switching devices (memristive systems,  Eqs.~(\ref{Geq1}) and (\ref{Geq2})). For this purpose, we suggest to use a capacitor-memristor circuit as sketched in Fig.~\ref{fig:2}(a). The circuit is  driven, in the simplest case, by a rectangular voltage pulse followed by zero bias (Fig. \ref{fig:2}(b)). We emphasize that any other waveform followed by zero bias, that would cause the tested device to switch, would provide similar results.

Our main idea is based on the following \textit{duality property of the capacitor-memristor circuit:} For {\it any} initial state of the memristor and {\it any} form of the applied voltage $V(t)$, the {\it final} state of the memristor (its resistance) {\it must} be identical to its {\it initial} state, if the capacitor charge finally returns to its initial value. In practice, for any particular device subject to this test, the return into the initial state should be verified in a reasonably wide range of applied voltage amplitudes and for various representative initial states of the tested device. The test is passed if the initial and final states (the memristor resistances) are within the limits of experimental accuracy.

We also stress that the presence of an additional parasitic capacitance (typically originating from the
parallel-plate geometry of the device electrodes~\cite{pershin11a}), would {\it not} invalidate the test we suggest. The reason is that also this capacitance will discharge by the end of the circuit evolution and, thus, the charge flowed through the memristor will be zero, when the final state of the memristor is read. Therefore, for an ideal memristor, its final resistance (which corresponds to the state with $q=0$) {\it must} coincide with its initial one {\it irrespective} of the presence of additional parasitic capacitances (if their final charges are the same as the initial ones). Likewise, our test is valid {\it irrespective} of the physical mechanism that gives rise to memory.

The duality property of the capacitor-memristor circuit we discussed above is self-evident. Indeed, since the current can not flow across the capacitor, the capacitor in the circuit of Fig. \ref{fig:2}(a) serves as a {\it charge-tracking} device so that the charge across the capacitor defines the instantaneous value of the memristance. In what follows we first exemplify the test by considering an ideal memristor and a threshold-type memristive system~\cite{pershin09b}. We will numerically model the test applied to the above mentioned devices and compare the test outcomes.

For the sake of definiteness, we assume that the ideal memristor is described by
\begin{equation}\label{eq:2}
  R_M(q)=R_0+\alpha \left(q-q_0 \right)^2,
\end{equation}
where $R_0$ is the minimal value of $R_M$, $\alpha$ is a constant, and $q_0$ is used to define the initial value of $R_M$. Note that any other model of ideal memristor would provide similar test results.

For a typical memristive system, we instead choose the model of threshold-type memristive elements (describing physically sound, experimentally
realized systems) as formulated in Ref.~\cite{pershin09b}:
\begin{eqnarray}
I(t)&=&R_M^{-1}V_M(t) \label{eq:3}
\\
\frac{\textnormal{d}R_M}{\textnormal{d}t}&=&
\begin{cases} \beta(V_M-V_{t}) \;\;\;\textnormal{if} \;\;\; V_{t}<V_M \\
\beta(V_M+V_{t}) \;\;\;\textnormal{if} \;\;\; V_M<-V_{t} \\
 0 \;\;\;\;\;\;\;\ \textnormal{otherwise} \end{cases} ,
\label{eq:4}
\end{eqnarray}
where the memristance $R_M$ is used as the internal state variable \cite{chua76a}, $\beta$ is the switching rate, $V_{t}$ is the (positive) threshold voltage. Moreover, it is assumed that the memristance is limited to the interval [$R_{on}$, $R_{off}$], where $R_{on}$ and $R_{off}$ are the low and high resistance states of the memristive system, respectively~\cite{pershin09b}.

\begin{figure*}[t]
\centering{(a)\includegraphics[width=80mm]{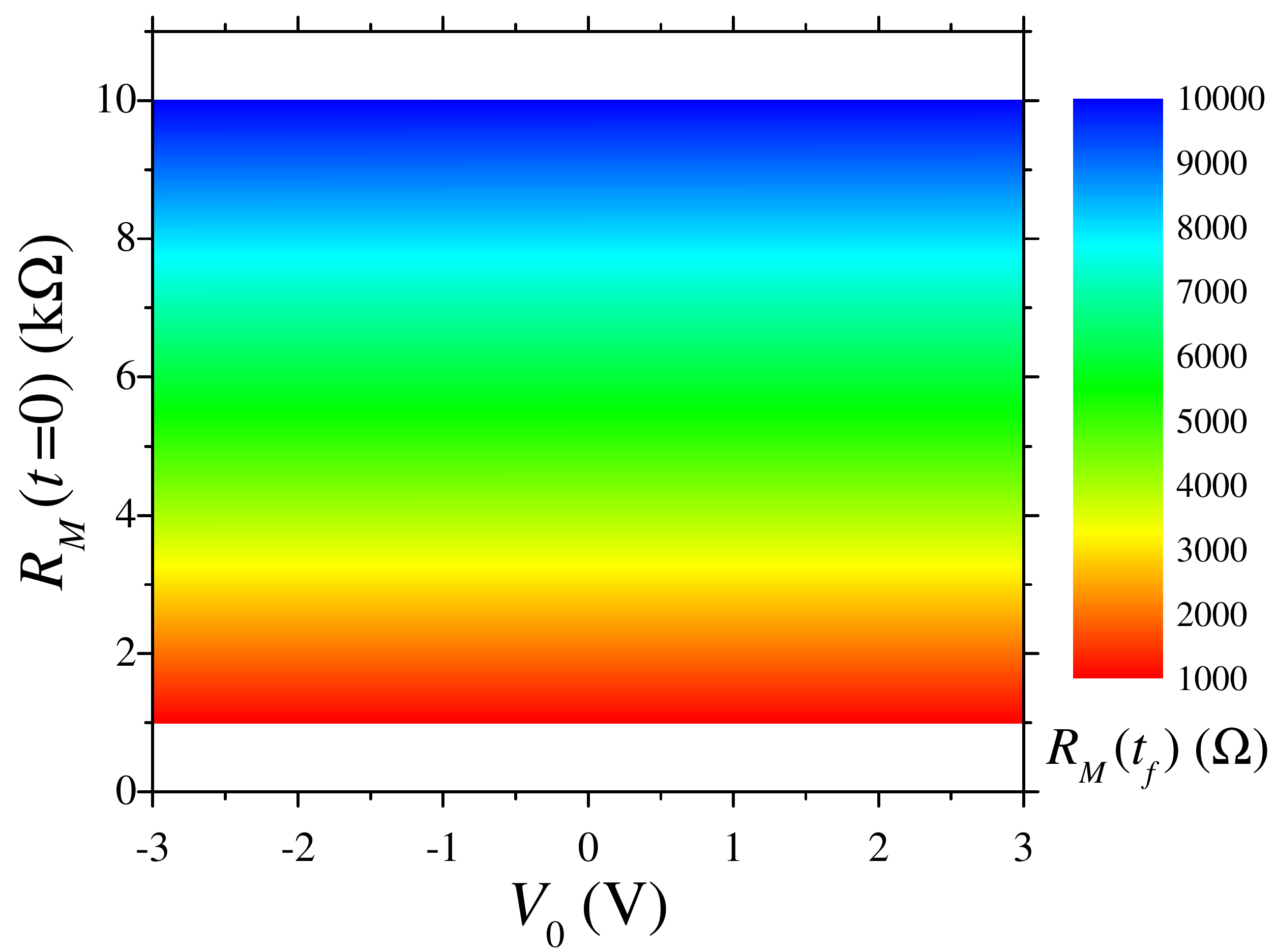}\;\;\;\;(b)\includegraphics[width=80mm]{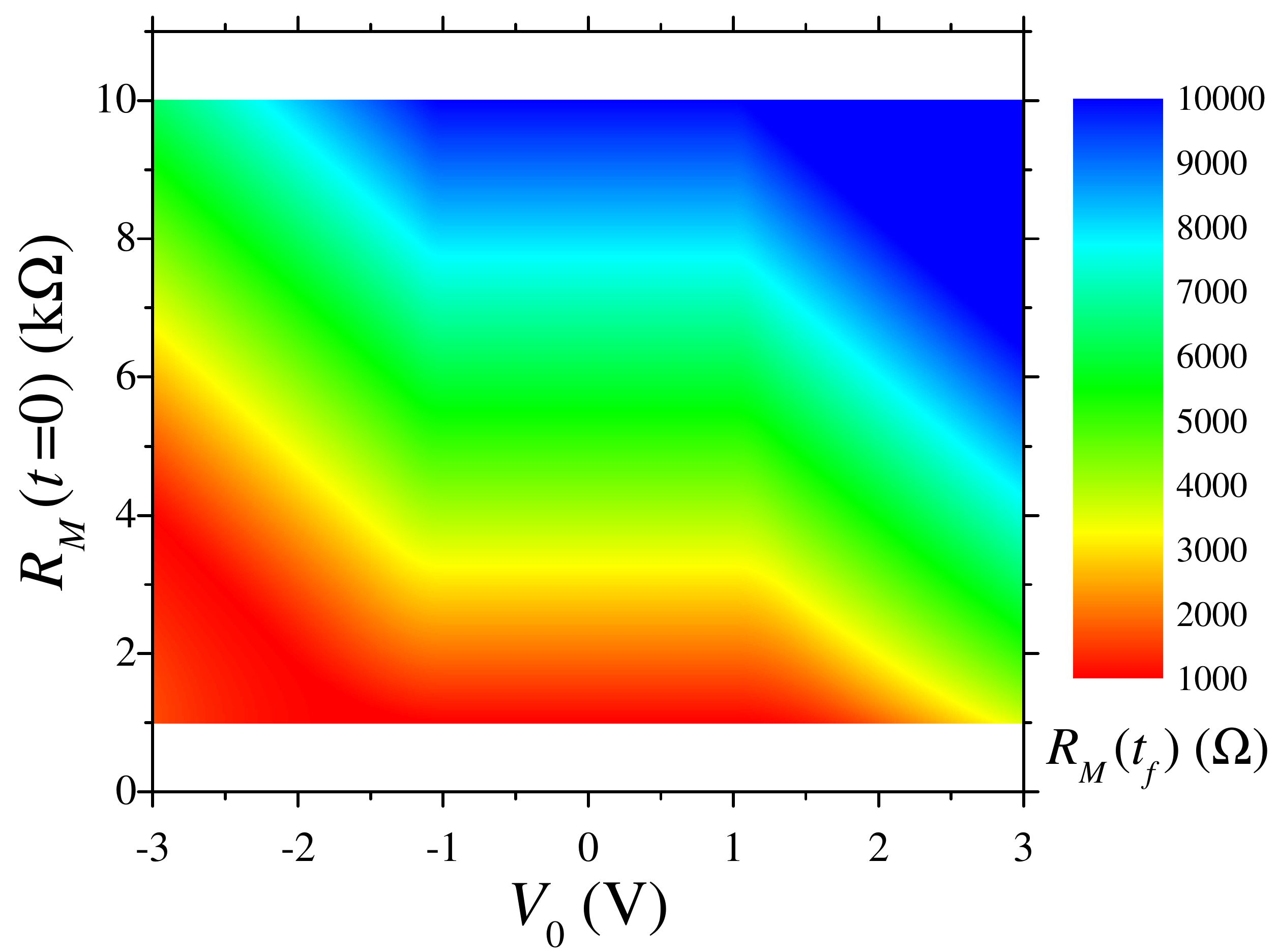}}
\caption{Map of the final states of (a) an ideal memristor and (b) a threshold-type memristive system as a function of their initial state, $R_M(t=0)$, and voltage pulse amplitude, $V_0$. Parameters used in these simulations are $R_0=R_{on}=1$~k$\Omega$,
$R_{off}=10R_{on}$, $V_t=1$~V, $C=2$~$\mu$F, $t_f=100$~ms and the voltage pulse width of 2~ms. The initial memristances, $R_M(t=0)$, range from $1$~k$\Omega$ to $10$~k$\Omega$, and the pulse amplitudes, $V_0$, range from $-3$~V to $3$~V. Colors  range from high values of final resistance (blue) to low values of final resistance (red). An ideal memristor shows the same value of final resistance as the initial one for the entire range of initial states and voltage amplitudes. Instead, a memristive element clearly shows a structure in the corresponding map.}
\label{fig:3}
\end{figure*}

In our numerical simulations, the circuit in Fig.~\ref{fig:2}(a) was modeled for different values of the initial memristance, $R_M(t=0)$, and pulse amplitude, $V_0$. To reach (at the final moment of time) the initially uncharged state of the capacitor, a sufficiently long circuit evolution time was selected. Fig. \ref{fig:3} presents the maps of the final states, $R_M(t_f)$, of an ideal memristor (Fig.~\ref{fig:3}(a)) and a threshold-type memristive system (Fig.~\ref{fig:3}(b)) as a function of their initial state, $R_M(t=0)$, ranging from $1$~k$\Omega$ to $10$~k$\Omega$, and voltage pulse amplitude, $V_0$, ranging from $-3$~V to $3$~V.

Figure \ref{fig:3}(a) clearly shows that, as expected, in the case of an ideal memristor, the final and initial states of the resistance {\it must} always be the same. Therefore, an ideal memristor would pass the test we propose. On the other hand, the final states of the threshold-type memristive system strongly depend on the initial state and pulse amplitude, $V_0$, and coincide with the initial states only when $-V_t<V<V_t$, namely, when the applied voltage is not sufficiently strong to switch the device resistance. Hence, the memristive system {\it does not pass} the test.

We emphasize that the plot in Fig. \ref{fig:3}(b) depends on the capacitance C, which must be properly selected as follows. Since our test evaluates the resistance switching characteristics of tested devices, the resistance switching must occur in wide regions of the parameter space. Therefore, with regard to the choice of $C$, the latter should be sufficiently large to provide a suitable voltage fall across the tested device for a sufficient interval of time to induce its switching. In other words, the $R_MC$ time must be sufficiently larger than (or on the same order of) the time it takes the device to switch.

\begin{figure}[bt]
\centering{\includegraphics[width=80mm]{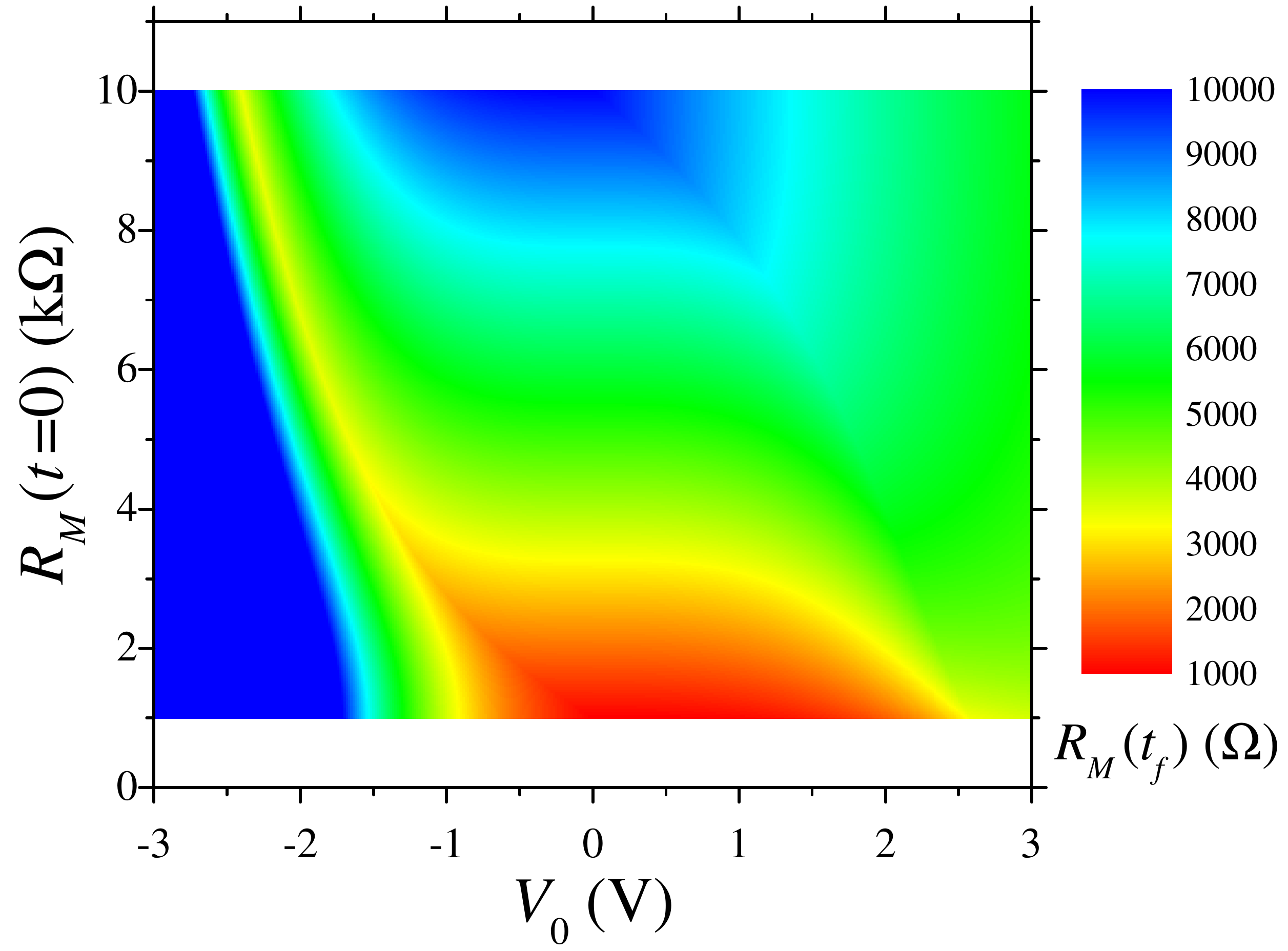}}
\caption{Map of the final states of a sinh-type memristive system. Model and calculation details are given in the text.}
\label{fig:4}
\end{figure}

To further illustrate our test, we plot the map of final states found for memristive systems described by a sinh-type model of switching~\cite{strukov09a}. Here, all the calculations have been performed similarly to the above with the difference that Eq. (\ref{eq:4}) is replaced with
\begin{equation}
  \frac{\textnormal{d}R_M}{\textnormal{d}t}=\gamma\;\textnormal{sinh} \left( \frac{V_M}{V_t} \right)  \label{eq:5},
\end{equation}
where $\gamma$ is a constant and $V_t$ is the threshold voltage. The map of final states obtained for $\gamma=10^3$~k$\Omega$/s and $V_t=1$~V is presented in Fig. \ref{fig:4}. Similarly to Fig. \ref{fig:3}~(b), the map contains a structure showing that Eq. (\ref{eq:5}) device is {\it not} an ideal memristor. This emphasizes again that only the devices described by $R(q)$ will pass the test introduced in this work.

In summary, since a pinched hysteresis loop is {\it not} a definitive test to determine whether any resistive switching device exists that can be represented by Eq.~(\ref{eq:1}) ({\it ideal} memristor), we propose a simple and yet {\it unequivocal} test that would accomplish this. Our test is based on the duality property of a capacitor-memristor circuit. This duality requires that for {\it any} initial state of the memristor (its inital resistance) and {\it any} form of the applied voltage $V(t)$, the {\it final} state of the memristor (its final resistance) {\it must} be identical to its initial state, if the capacitor charge finally returns to its initial value. In practice, a wide enough range of initial states of the memristor and forms of the applied voltage would suffice to render this test valid.

At the moment, we are not aware of any experimental resistance switching device that can pass our test. We thus hope the test we
propose will be used in all those devices that are currently classified as ideal memristors to check if that classification is indeed correct or simply serves as a rough mathematical approximation to physical systems that are actually more complex than Eq.~(\ref{eq:1}) would suggest.

\section*{References}

\bibliographystyle{IEEEtran}
\bibliography{memcapacitor}

\end{document}